# Discourse Analysis of Covid-19 in Persian Twitter Social Networks Using Graph Mining and Natural Language Processing

Omid Shokrollahi[1], Niloofar Hashemi[2], Mohammad Dehghani[3]

**Amirkabir University of Technology, Tehran , Iran**

**Allame Tabatabaei Universiy, Tehran ,Iran**

**Tarbiat Modarres University, Tehran, Iran**

## Abstract

One of the new scientific ways of understanding discourse dynamics is analyzing the public data of social networks. This research's aim is Post-structuralist Discourse Analysis (PDA) of Covid-19 phenomenon (inspired by Laclau and Mouffe's Discourse Theory) by using Intelligent Data Mining for Persian Society. The examined big data is five million tweets from 160,000 users of the Persian Twitter network to compare two discourses. Besides analyzing the tweet texts individually, a social network graph database has been created based on retweets relationships. Extending on the previous work, which has emphasized using centrality metrics to analyze social network influencers, we use the VoteRank algorithm to introduce and rank people whose posts become word of mouth, provided that the total information spreading scope is maximized over the network. These users are also clustered according to their word usage pattern (the Gaussian Mixture Model is used). The constructed discourse of influential spreaders is compared to the most active users. Then the changes of individual discourse have been evaluated for these two subpopulations over time, corresponding to this period's most important news. This analysis is done based on Covid-related posts over eight episodes from the end of January to mid-May -2020. Also, by relying on the statistical content analysis and polarity of tweet words, discourse analysis is done for the whole mentioned subpopulations, especially for the top individuals. The most important result of this research is that the Twitter subjects' discourse construction is government-based rather than community-based. The analyzed Iranian society does not consider itself responsible for the Covid-19 wicked problem, does not believe in participation, and expects the government to solve all problems. The most active and most influential users' similarity is that political, national, and critical discourse construction is the predominant one. However, the contrast is that the more scientific view (Medical Discourse) with the positive tonality is predominant for influential individuals. In addition to the advantages of its research methodology, it is necessary to pay attention to the study's limitations. Suggestion for future encounters of Iranian society with similar crises is given.

## Discourse Analysis:

This research has a Post-structuralist qualitative exploratory nature. It's main method is discourse analysis which it means that our access to reality is always through language. In fact, through language, we create representations of reality. Language isn't a reflection of a pre-existing reality; in fact, it plays a role in constructing reality. This does not mean that there isn't reality, meanings and representations are real things, but they find meaning only through discourse.

Structuralist linguistics owes much to Saussure's ideas that language is a system that is not dependent on the reality that reveals it. Saussure attributes signs to form (signifier) and content (signified) with arbitrary relationship between them. Therefore, the meaning of words is not inherent in them and is based on social conventions that we attribute special meaning to certain sounds.

But Post-structuralism begins with this idea of structuralism that signs take meaning by the internal relations within the network of signs, not by their relation to reality. But Poststructuralists like structuralists, do not recognize language as an inflexible, fixed, and pervasive structure. In the Saussurean tradition, the structure of the language is like a fishing net, in which each sign is a node with a definite position; Thus, structuralism is based on the assumption that each sign has a specific place in the network and a fixed relationship with other signs, so its meaning is fixed. But in the poststructuralist belief, although signs derive their meaning from differences with other signs, but this relationship is no longer a fixed one.

The concept of discourse in Laclau and Mouffe's theory holds that all objects and actions are meaningful and that their meaning is formed by referring to specific systems of meaningful differences. Thus, the field of discourse is not limited to purely linguistic phenomena, and objects never have an extra-discursive meaning.

Articulation is the creation of nodal points that stabilize meaning to some extent, but this consolidation of meaning is always incomplete due to social openness, and this incompleteness is the product of the constant overflow of any discourse, due to the infinity of field of discursivity. Specific discourses establish a part of social meaning, but the characteristic of discourse fields is the "surplus of meaning", which cannot be emptied by any particular discourse.

Discourse requires extra-discourse and can never fully articulate the elements of discourse, because there are forces that are defined against discourse.

The central signifier is a concept around which other signifiers are articulated, and the floating signifier meaning is unstable and numerous; Thus, different groups compete with each other for the appointment of their intended signified. A micro-network is some of the structure of a discourse network that belongs to several major discourse signifiers that establish a definite relationship with each other.

Thus, the main signifiers in the whole discourse space and the special signifiers in the micro-networks have meaning. Equivalent signifiers have similar conditions in terms of the accumulated power in the discourse network space.

## Methodology:

In this research, we study the problem of post-structuralism discourse analysis of the Corona phenomenon (with Laclau & Mouffe's method) by intelligent data mining and content analysis[1].

Goal: how is the advent of the Corona phenomenon constructed in Persian society?

---

[1] Appendix A

The central question: has Corona phenomenon been constructed in a government-based fashion (requires to be responded by government, a top-down view) or instead it has been constructed in a community-based fashion (requires to be responded by the community, a bottom-up view)

Subsidiary questions:

1. What are the dimensions of government-based discourse and community-based discourse?

2. How was the evolution of the sign of corona in the time interval of the study?

3. How news spread affected the transformation and transfiguration of the discourse space?

4. How different subjects (human agency) affected transformations and transfiguration of the discourse space?

5. How the Corona phenomenon is constructed in Persian society help to confront it properly?

Steps:

1-Exploring and processing the raw data

2-Creating a 100-word dictionary for each discourse of study by studying the most used hashtags

3-Classifying the discourses

4-Determining the time intervals as episodes upon analyzing important news.

5-Examining the role of subjects (human agency) in constructing discourse.

6- Sociological analysis

## Results:

### 1. The most active users:

In the next figure (Fig. 2), we demonstrate the evolution of the discourse space for 50 individuals of the most active users, people whom more used hashtags in the network, from different perspectives. This diagram shows the intensity of dictionary words related to government-based discourse compared to the community-based discourse's dictionary. It is evident that the most active individuals who used words of our dictionary intensively constructed a government-based construction for the Covid-19. Furthermore, the special signifier of their discourse network belongs to the government-based dictionary. The climax for the usage of the community-based dictionary occurs in episodes 3 and 4, and a relatively smaller number of people has used this. These two episodes correspond to the time when the Iranian students in china have returned, the pandemic was first announced, and the national headquarters established. On the other hand, the usage of the words from the government-based dictionary had increased after episode 4 when the pandemic was widely announced; this reveals that the discourse space is significantly political.

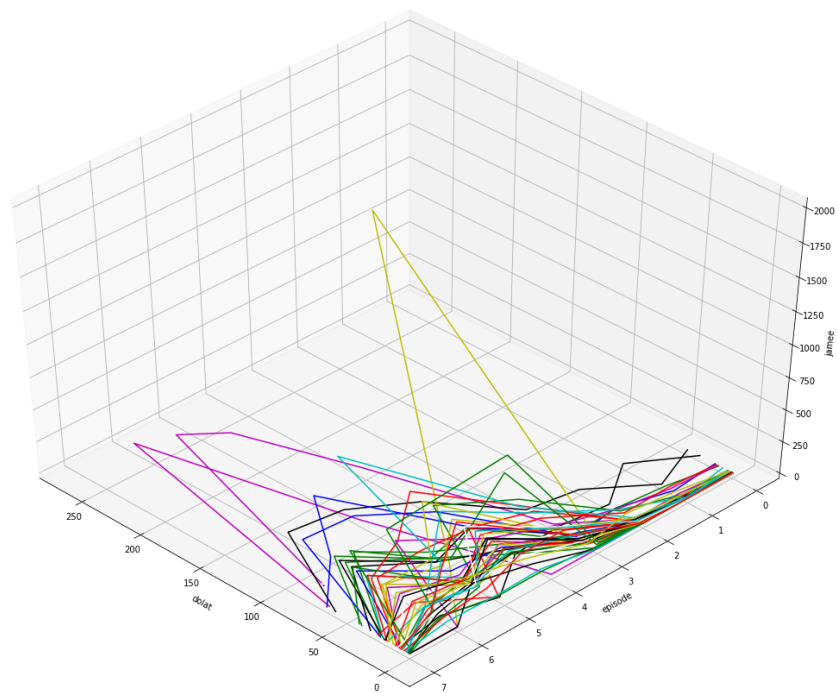

**Fig. 2: The evolution of the discourse space for 50 individuals of the most active users**

Investigating the diagrams for the most tweeted hashtags also proves the previous statement of analysis. Among the ten topmost active users, only one person used more community-based hashtags and the others, in most episodes, typically use more government-based hashtags than community-based hashtags.

However, the diagrams of the used hashtags for the top 10 ranks of community-based discourse activists who used the community-based dictionary words are worth considering. Except for one, nine of them created community-based hashtags, mostly in episode 4, the Corona pandemic announcement. It seems that afterward, they have ceased to have such an attitude. This fact brings to mind many reasons, including the arousal of emotions when a confrontation with a national crisis, being influenced by other subjects or individuals or politicizing the discourse space according to the government's overall performance.

## 2. The most influential people

From the analysis of 5 influential people, it could be concluded that although a completely medical signified is constructed for Covid-19, but medicine is also a micro-network in the political space and becomes meaningful in the existing national power relations.

Therefore, it seems that the main difference between the most influential people on Twitter and the most active ones is that the medical construction is more effective. The similarity is a master of political construction, both in the national sphere and in the critique of existing power relations.

But in the continuation of the analysis of the discourse space of the most influential people on Twitter, according to the figure 3, it is clear that in the discourse network of these people, the signifiers of the government-based micro-network are the main special signifiers. It could be concluded that the political construction, attributed to power relations is dependent on government and the top-down intervention of Corona is superior in relation to a participatory model, social issues, and empathetic and bottom-up approach.

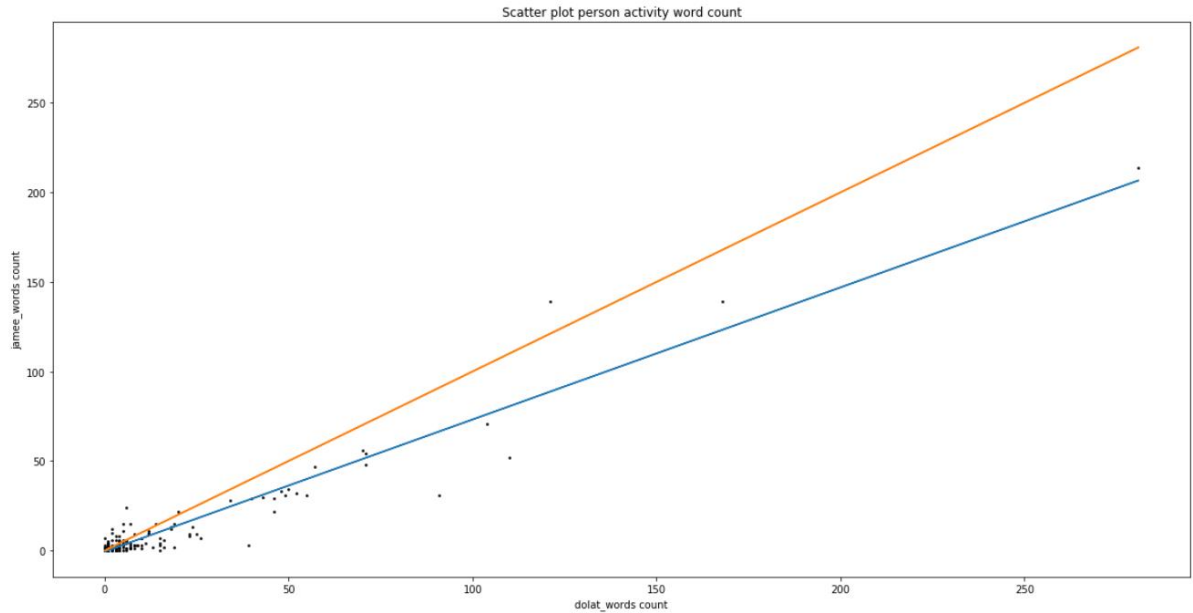

Fig. 3: Scatter plot person activity word count

But the evolution of the discourse space in the next figure (Fig. 4) is well visible. It is clear that the most influential people on Twitter have a much more government-based construction of Covid-19, and the special signifier of their discourse network are signifiers of a government-based dictionary. Thus, the dominance of power-attributed construction in front of the Corona issue is the most obvious result of the Corona discourse analysis in Iranian society in the Twitter community.

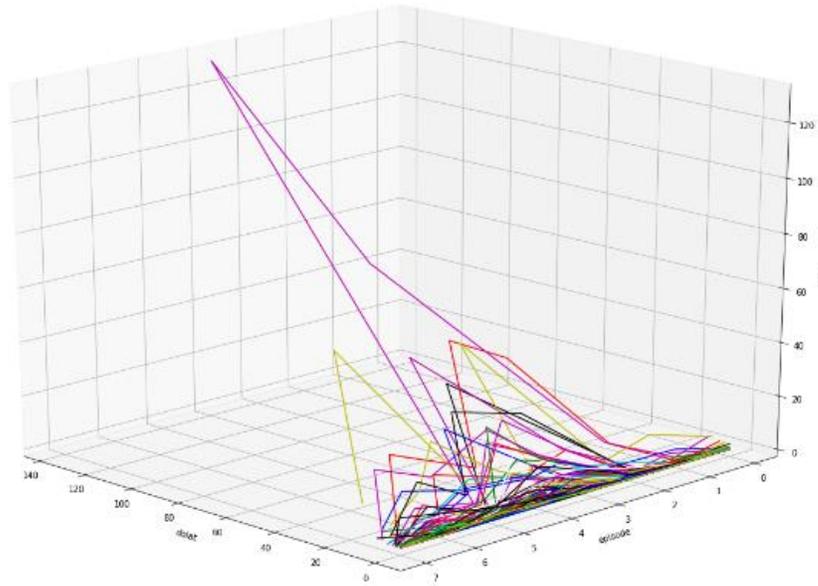

**Fig. 4: The evolution of the discourse space for 50 individuals of the most influential users**

But another aspect of this pattern illustrates the fact that in episodes 3 and 4, the most influential people on Twitter had a more community-based construction than other time periods. These episodes are before the news of the spread of the virus in Iran and the first days of exposure to it, that is, when there was little opportunity for analysis, Covid-19 created more social issue and it discontinued because of frustration and a more political view of the issue.

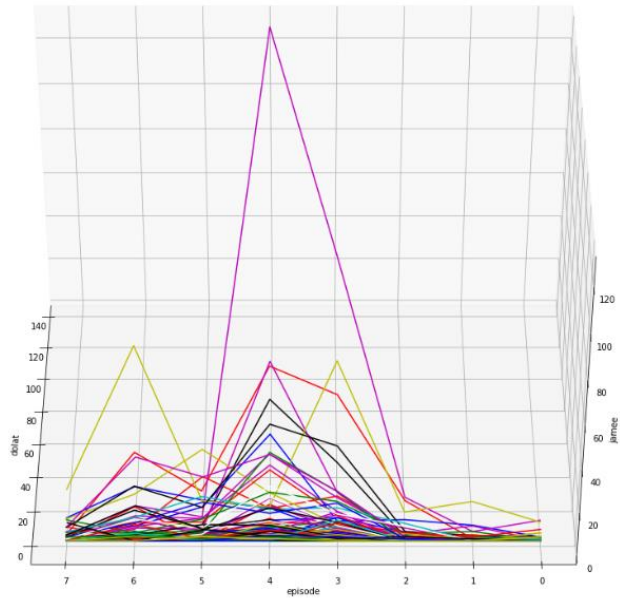

**Fig. 4: The evolution of the discourse space for 50 individuals of the most influential users**

Finally, the results of clustering confirm the dominance of government-based discourse over community-based discourse in the most influential people.

## Conclusion

1. The most important result of this research is that the Twitter subjects have constructed much more government-based discourse more than the community-based one. Therefore, it seems that the participatory perception of phenomena is an absent signifier of the discourse of Iranian society. The analyzed Iranian society does not consider itself responsible for the phenomena, does not believe in participation, and the society is waiting for the solution of all issues by the government. Of course, the bias of the obtained results is quite clear due to the limited population studied in this section. What was observed in the early days also had many noteworthy participatory dimensions. These people are not necessarily Twitter activists and influencers, and may have spent most of their time in various forms of participation!

2. The similarity between the most active and influential people is the political, national and critical construction, but the difference is in the more scientific view (medical discourse) by the positive tone of the influencers.

3. Regarding the dynamic image of the mentality of the society and the discourse of the Iranian society, the discourse evolution in both groups, in the episodes about announcing the outbreak of Corona in Iran, has the most community-based level, which shows the feeling of empathy in the moment of crisis. But in the most active people, government-based has increased in the final episodes, which has not changed much in the most influential people. Thus, the most active people in the face of governance approaches have gradually taken a more critical view. But the most influential people have a relatively stable trend of criticism.

4. The mentality of the Iranian society in the coming months couldn't be shown, because there is no data available during the corona issue consolidation period and all data belong to the time before the outbreak or the first time intervals after it.

5. The only model that could be predicted in the future confrontation of Iranian society with similar crises is the empathetic reaction in the early moments, which, if viewed as social capital by the regime, could be considered as profitable source for solving crisis and reduced the political critical side of society and turned it into an energetic activist force, which ultimately reduced the conflict and the nation-state gap.

6. Finally, it seems that instead of focusing on negative news and international failures in resolving the Corona crisis, it is possible to focus on the participatory and responsible dimensions of nations in the news, etc., so that this aspect of participation should also be strengthened in Iranian society.

**Acknowledgements:**

We would like to thank Iran Cognitive Sciences and Technologies Council

# Appendix A

## Social network analysis:

As a growing new field, there are different problems under the umbrella of social network analysis; one is exploring the network's patterns to extract knowledge about human behavioral patterns and obtain valuable discoverable information about human interactions.

The Twitter social network is a dynamic graph changing over time. This experiment will consider this graph's evolution as a sequence of short snapshots on the network's state. For this purpose, we adopted eight-time intervals to correspond to the real-world occurrences by analyzing essential news and events. These reflect the critical political decisions or events during the Corona pandemic's advent and determine the network's state.

Our model for investigating activity on Twitter is a weighted multi-layer network for which we represent every discourse of study as a separate layer. We define post-retweet in each time interval between individuals as edges among nodes that stand for user accounts to organize the network structure. The retweet network for investigating information diffusion mechanism has been studied by other authors [1]. The content of the posts is then analyzed to detect the type of discourse or edge types. The number and type of retweeted posts determine the edge weights. Therefore we have a directed graph that Wij or the edge weight between node i and j indicates the number of tweets between the author with id I and the person who retweets it with id j. due to data limit restrictions, we call off the likes and comments.

We formulate two ways to analyze influencers. The first one considers the single tweet structure and the second one also examines the underlying network structure. In the latter one, we exclude the tweets that do not incite others to retweet its text. However, we inspect these posts in the framework of the former analysis. Their hashtags, similar to every other correspondence, will be extracted for user activity analysis; they also indicate the top influencers at each time interval.

To identify the influential spreaders, we built a network of all activities aggregated across all episodes and network layering regarding the post's discourse. For this purpose, weighted edges show the number of retweeted posts that one makes of another individual.

The connection between nodes in a network may inform us about the node's importance. For instance, we can identify which node is most influential in spreading information in a social network. For calculating nodes' importance, we may use different quantitative indexes known as centrality measures. Degree distribution, betweenness, and closeness may be useful to study influencers of the social network. Degree centrality is the number of links incident upon a node (i.e., the number of ties a node has). Indegree counts the number of connections directed to the node (head endpoints), and outdegree is the number of relations that the node refers to others (tail endpoints). So it may indicate how many people are influenced by a node or how many different people may influence one person in a given social situation. A node with higher betweenness centrality would have more control over the network because potentially more information will pass through that node. Betweenness indicates to what extent edges and people are essential for group coherence.  Closeness centrality is a measure of the average shortest distance from each vertex to each other vertex. It is a useful measure that estimates how fast the flow of information would be through a given node to other nodes.

Recognizing components or clustering social network for detecting communities of interdependent humans is another approach of network analysis. People who have more communication with each other are said to be strong-tie and classified in the same communities. On the other hand, people who do not belong to the same community do not have strong connections, or at least it's what the data says. Strong-tie edges are often readily obtained from the social network as users often participate in multiple overlapping networks via features such as following and retweeting[2]. We may investigate whether people who participate in the same communities use identical discourses. Clustering is an optimization problem, and different objective functions and, therefore, algorithms are ben defined for this purpose. For example, the stochastic block model(SBM) measures the required information to explain the data based on statistical evidence to obtain the most simple model with the least information entropy to explain the grouping[3,4,5]. It has a high generalization power based on observed data and proposes predictions beyond data. This ability becomes crucial since we may have missing edges or inaccurate ones.

We built a big network consisting of more than a million links and 160 thousand nodes. As the size of the network grows, visualizations that serve for summarization and gaining insight become impractical. In the next section, we describe how we select a sample population as decentralized influential spreaders. The underlying mechanism, known as the Vote-Rank algorithm [6], is selective and not random and, above all, somehow serves to analyze the spread content at the whole population level.

## Selecting the sample population using the vote-rank algorithm

Members of the social network dispute different ideas and spread them among their followers to spread it among diverse audiences. Users' activities and exchanging messages and word of mouth in a social network can be a particular case of the "effective information spreading" problem. Maximizing the scale of spreading is the target. We define the question of choosing initial nodes as source spreaders to achieve a maximum spread as an influence maximization problem.

It is a simple yet effective iterative method to choose a set of influential spreaders. They are elected one by one according to their voting scores obtained from their neighbors. At each iteration, the elected spreader's voting ability will be set to zero, while that of its neighbors will be decreased by a factor. The strategy is to choose top-r ranked nodes as spreaders according to influence ranking methods such as PageRank, Cluster Rank, and k-shell decomposition.

However, other approaches suffer from the possibility that some spreaders are so close together that they overlap the sphere of influence. But in the vote-rank [6], the selection probability of its neighbors and neighbors' neighbors will decrease. Under this mechanism, the selected nodes are far apart and are essential in their local structure ad have high connections with neighboring nodes. Therefore the selected candidates are expected to resemble diversity.

In summary, by utilizing information of r − 1 ranked nodes to rank the rth node, we get an obvious boost on information spreading in complex networks, especially in large scale networks.

We do not discuss the optimal value for r key spreaders. But since we will have a qualitative study, the size of this population is chosen relatively small.

After selecting an influencing population with the prescribed mechanism, we expect to explore posts and discourses with the maximum spread on the network as the sample population.

First, we explore the generated word-cloud for each user and do a qualitative study on the words for some users; then, we cluster all the selected users using the frequency pattern of the used words in their posts. After recognizing the discourse that these words belong to, we can detect different voices that contribute to each discourse by clustering these users. Each of these voices may equally stem from a unique Gaussian component. It is trivial that the data dimensions (number of words belonging to each discourse's dictionary scheme) do not follow a homogenous pattern among different subpopulations. They may show various proportions; therefore, it's not convenient to use the k-means algorithm for such clustering.

## Clustering users with Gaussian mixture models:

Gaussian mixture models [7] are a probabilistic model for representing normally distributed subpopulations within an overall population. Since subpopulation assignment is not known, this constitutes a form of unsupervised learning. It is a superposition of multiple Gaussian distributions. GMM assigns a two-dimensional normal distribution to each data point( where the number of dimensions is the number of discourses understudy) with appropriate mean and variance to maximize the posterior probability of the observed data. This formula is a weighted mean for K (number of clusters or subpopulations) Gaussian distribution. Weights differ proportionally to the number of samples for each component.

Now that we have our basic parameters, we will calculate the probability that our distribution for each data point generates. Thus the points which belong to the same cluster will have high probabilities & the outliers will have low probabilities.

When the number of components K is not known a priori, it is typical to guess the number of components and fit that model to the data using the EM algorithm. This model fitting is done for many different values of K. Usually, the model with the best trade-off between fit and number of components (simpler models have fewer components) is kept. Bayesian Information Criterion (BIC) scores each model to select among a finite set of models; the following comparison evaluates models with different components with different assumptions on their covariance matrices.

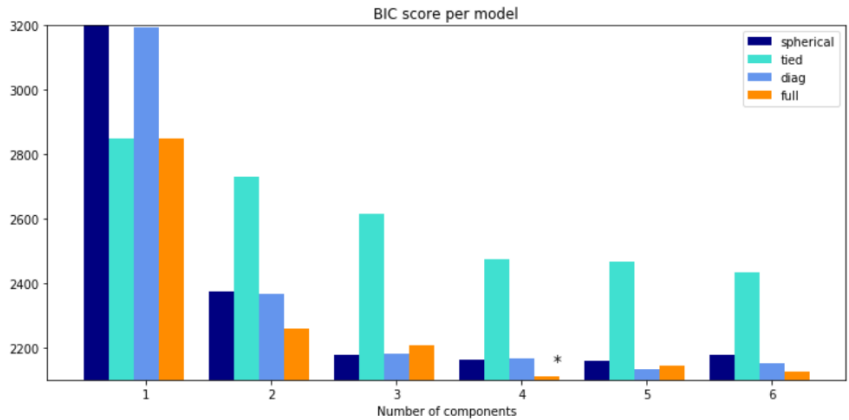

**Fig. 1: comparing BIC score for different Gaussian Mixture models, to select best model with minimum BIC score**

## Content Analysis:

We inspected all the data and studied a sorted list of hashtags and the number they have appeared. We reviewed 64% of all the used hashtags in all posts with an eye examination. Then we adopted two sets of 100 words from this list for studying a scheme of the two discourses.

In this experiment, we used a Rule-based algorithm to identify the discourses from the tweets' words. Suggestions for a deep learning-based approach for discourse analysis could be future works.

We carried out a qualitative study on all of their posts' word clouds across different time episodes for some selected influencers. A word cloud is an image composed of words used in which the size of words indicates their importance or frequency. We used Word cloud to visualize the text data and show which words appeared more frequently in the user's posts collection at different time episodes. In this analysis, we discarded terms of no importance (stop-words) from the pack.

Finally, we examined the sentiment of the generated posts collection for each selected individual again using a rule-based method. To detect the polarity (from positive/ negative/ neutral category) of an individual's voice, we used a Persian words polarity dataset to count the number of words that belong to each class. We explored the collection of all their used words in tweets to evaluate the tonality of each user.